\documentclass[aps,pre,twocolumn,superscriptaddress]{revtex4}
\usepackage{graphicx}
\usepackage{epsfig}
\begin{document}

\title{Unraveling DNA tori under tension}

\author{C.\ Battle}
\altaffiliation[Present address: ]{Drittes Physikalisches Institut, Georg-August-Universit\"{a}t, 37077 G\"{o}ttingen, Germany}
\affiliation{Institute for Theoretical Physics, Universiteit van Amsterdam, 1018 XE Amsterdam, The Netherlands}

\author{B.\ van den Broek}
\altaffiliation[Present address: ]{Leiden Institute of Physics, Leiden University, 2333 CA Leiden, The Netherlands}
\affiliation{Department of Physics and Astronomy and Laser Center, Vrije Universiteit, 1081 HV Amsterdam, The Netherlands}

\author{M.C.\ Noom}
\affiliation{Department of Physics and Astronomy and Laser Center, Vrije Universiteit, 1081 HV Amsterdam, The Netherlands}

\author{J.\ van Mameren}
\affiliation{Department of Physics and Astronomy and Laser Center, Vrije Universiteit, 1081 HV Amsterdam, The Netherlands}

\author{G.J.L.\ Wuite}
\affiliation{Department of Physics and Astronomy and Laser Center, Vrije Universiteit, 1081 HV Amsterdam, The Netherlands}

\author{F.C.\ MacKintosh}
\email{fcm@nat.vu.nl} 
\affiliation{Department of Physics and Astronomy and Laser Center, Vrije Universiteit, 1081 HV Amsterdam, The Netherlands}

\date{\today}

\begin{abstract}
Motivated by recent experiments, we develop a model for DNA toroids
under external tension.  We find that tori are the equilibrium
states for our model up to a critical tension, above which they
become only metstable.  Above this tension, we find a cascade
of transitions between discrete toroid states that successively lowers the
winding number, until the ground state (rod) is reached. In this
process, this model predicts a nearly constant force plateau as a function of extension, 
in agreement with experiment.
\end{abstract}

%\pacs{87.16.Ka,62.20.Dc,82.35.Pq}
%\keywords{bla}

\maketitle \section{Introduction}

It has long been recognized that the conformation of polymer chains
depends on the solvent properties of the
environment~\cite{deGennesScaling,Doi-Edwards}.  In particular, polymers in poor
solvent conditions effectively attract each other in an attempt to
exclude the solvent, forming collapsed structures that minimize
surface contact with the solvent.  For flexible polymers, this leads
to compact globules of roughly spherical shape, whose kinetic
pathway has been shown to involve the formation of a pearl necklace
and gradual diffusion of large pearls to the chain
end~\cite{Ostovsky1994,Chu1995,Buguin1996}.

In the case of semiflexible polymers such as DNA, which exhibit a
substantial bending stiffness, the energetic penalty for bending
causes spherical globules to be energetically disfavored.  The
apparent equilibrium states for these polymers have been shown to be
rings or toroids~\cite{PhysToday2000}, as these structures balance
the tendency for the polymer to condense due to effective
polymer-polymer attraction with the tendency to minimize curvature
due to bending stiffness.  These condensed states have been studied
theoretically~\cite{grosberg1979, bloomfield1991, Ubbink1995,bloomfield1997,
vasil1997, pereira2000,Schnurr}, observed in
experiments~\cite{lifanding1992,
fang1998surf, shen2000, martin2000, liu2001interm, hansma2001surf}, and shown by computer
simulation~\cite{noguchi1996, byrne1998kin,
noguchi2000bd,Schnurr2000,stevens2001}.

\begin{figure}[h]
\centering \includegraphics[width=8cm]{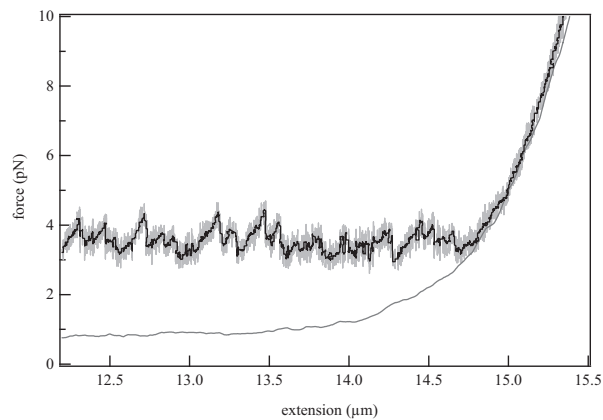} \caption{
Measured force-extension curve on condensed DNA.  A lambda-DNA molecule (48502 base pairs, or $\simeq16.4\mu$m in length) was stretched using optical tweezers in the presence of spermine4+ (black line, averaged to 8 Hz, raw data at 128 Hz shown in light gray). In contrast to the force-extension curve of uncondensed DNA (solid gray line) we find a force plateau of approximately 4 pN that persists throughout the curve. In this plateau small steps can be discerned, signifying unwinding of loops from a toroidal DNA condensate.}
\label{exptfig}
\end{figure}

In an effort to understand the dynamics of toroid formation, recent
experiments have explored the condensation of DNA under tension
\cite{Baumann2000,Murayama2003,Besteman2007,BramThesis,Bram}.  
Motivated by these experiments we analyze
theoretically a hierarchy of tori states and explore their
equilibrium and metastable structure under tension, as well as
transitions between toroid states. We find a sequence of metastable
tori under tension. Furthermore, we find that for winding numbers
larger than approximately 10, a nearly constant force plateau
emerges, which agrees well with recent observations, as illustrated in Fig.\ \ref{exptfig}.

In Sec.\ \ref{theModel} of this paper, we first define a simple,
non-thermal model that incorporates the essential physical effects
believed to give rise to DNA toroids: (1) the bending rigidity and
(2) the effective attractive interactions between DNA segments, such
as can arise in the presence of multivalent ions \cite{PhysToday2000}.
We summarize the relevant experiments in Sec.\ \ref{secExpts}. 
We then study the equilibrium and metastable states of this model,
as well as transitions among these states in Sec.\ \ref{theResults}.  
We conclude with a discussion of the implications of this model, and
the relationship of our results to the experiments.

\section{The Model}\label{theModel}

The first step in examining the equilibrium and metastable structure
of semiflexible polymer condensates under tension is to identify and
calculate their energy.  We model the conformational energy of a
toroid as in Schnuur, et al.~\cite{Schnurr}, where we assume
integer winding number toroids with a single radius of curvature at
zero temperature.  Our calculations describe
a simplified model of tightly packed filaments of vanishing
thickness.  We do not take into account any winding defects due to
topological constraints or variations in curvature due to filament
thickness.

We write the Hamiltonian of our system as a sum of bending and interaction terms
\begin{equation}
 H=H_{bend}+H_{int},
\label{hamiltonian}
\end{equation}
where the bending term models the energy of the curvature of the
major radius of the torus and the interaction term models the
self-attraction of the polymer, or equivalently its poor solvent
environment. The bending term can be straightforwardly calculated,
as the energy of the idealized chain described is simply that of a
series of circular rings, given by
\begin{equation}
 H_{bend}=\frac{\kappa}{2}\int_{0}^L ds~C^2(s)=\frac{\kappa}{2}\frac{L}{R^2}=\kappa \frac{2\pi^2 N^2}{L},
\label{bendingham}
\end{equation}
where $C(s)$ is the curvature, $L$ is the total filament length in the torus, $R$
is the torus radius, $N$ is the torus winding number, and $\kappa$
is a bending stiffness constant.  We note that since we consider
only integer winding number tori, $L=2\pi NR$, so we can
write~(\ref{bendingham}) in terms of a single quantity, $L$.

\begin{figure}[b]
\centering \includegraphics[width=7cm]{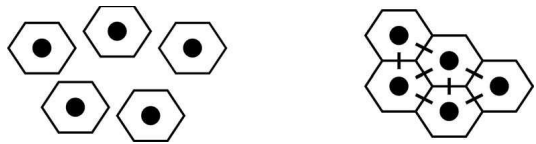} \caption{Sketch of cross section of $N=5$
torus, with hexagonal filament cross section. The perpendicular
lines represent bond between two sites on neighboring filaments. }
\label{bundle}
\end{figure}

For the interaction term, we assume a dense structure, in which
filaments pack tightly in their plane perpendicular to their local
axis. This suggests a simple, hexagonal packing of the filaments.
With such tight packing, we assume that the interactions are only of
the nearest neighbor type. In this limit, the filament can be
thought of as having six possible binding sites per unit length,
which can either form a DNA-DNA bond with another section of the
filament or can be exposed to solvent.  Bundling occurs when the
attractive interactions are sufficiently strong. In order to calculate the interaction term,
we define the number of bonds per cross section to be
$n_b=\frac{6N-n_s}{2}$, where $n_s$ is the number of solvent-exposed
sites.  We divide by two to avoid double counting, as a
DNA-DNA bond is equivalent to the merging of two binding
sites on neighboring filaments. Our interaction term can thus be
written as
\begin{eqnarray}
 H_{int}&=&-\gamma \int_{0}^{2\pi R} ds~n_b(s) \\ \nonumber
&=&-3\gamma L + \frac{\gamma}{2}\int_{0}^{2\pi R} ds~n_s(s) \\
\nonumber &=& -\alpha_1 \gamma L + \gamma \frac{\alpha_N}{N} L,
\label{intenergy}
\end{eqnarray}
where $\gamma$ is a surface tension parameter that characterizes the energetic cost of 
of solvent-exposed DNA and $\alpha_N$ is the
so-called coordination number~\cite{Schnurr}, which is equal to half the number of
solvent exposed sites per unit length along the torus circumference.  
We use the coordination number to enumerate
these sites, and it can be found by subtracting the number of
filament-filament bonds from $3N$.  As an example of this scheme,
consider the cases $N=5$ and 10:  for five filaments there are seven
bonds, resulting in a coordination number of 8 (see
Fig.~\ref{bundle}), while for ten filaments there are 19 bonds,
resulting in a coordination number of 11.  We then multiply the
coordination number by the interaction parameter, $\gamma$, to
obtain the surface energy per unit length of a bundle. Table~\ref{alphatable} lists
the first 24 coordination numbers. 
We replace the $3\gamma L$ term in the second line
of Eq.~\ref{intenergy} with $\gamma L\alpha_1$ to emphasize the physical
meaning of this term. The $\gamma  \alpha_1 L$ term comes from the
difference in surface energy between $N$ strands of unbundled
filament and $N$ bundled strands, and it reflects the physical
tendency of the torus to minimize unsatisfied bonds through
bundling.
%, with coordination numbers corresponding to ``magic'' winding numbers (to be discussed) in boldface.
%
\begin{table}
\begin{center}
\begin{tabular}{||r|r||r|r||r|r||}
\hline
\multicolumn{6}{|c|}{Table~1: Coordination Numbers} \\
\hline
$\quad\alpha_1$ & \quad 3 & $\quad\alpha_9$ & \quad 11 & $\quad\alpha_{17}$ & \quad 15 \\
$\alpha_2$ & 5 & \quad\textbf{$\alpha_{10}$} & \textbf{11} & \textbf{$\alpha_{18}$} & \textbf{15 }\\
$\alpha_3$ & 6 & $\alpha_{11}$ & 12 & \textbf{$\alpha_{19}$} & \textbf{15} \\
$\alpha_4$ & 7 & \textbf{$\alpha_{12}$} & \textbf{12} & $\alpha_{20}$ & 16 \\
$\alpha_5$ & 8 & $\alpha_{13}$ & 13 & \textbf{$\alpha_{21}$} & \textbf{16} \\
$\alpha_6$ & 9 & \textbf{$\alpha_{14}$} & \textbf{13} & $\alpha_{22}$ & 17 \\
\textbf{$\alpha_7$} & \textbf{9} & $\alpha_{15}$ & 14 & \textbf{$\alpha_{23}$} & \textbf{17} \\
$\alpha_8$ & \quad 10 & \textbf{$\alpha_{16}$} & \textbf{14 }& \textbf{$\alpha_{24}$} & \textbf{17} \\
\hline
\end{tabular}
\end{center}
\caption{Coordination numbers for tori winding numbers 1-24, with magic numbers in boldface.}\label{alphatable}
\end{table}
%\vspace{0.5 cm} 
%

\begin{figure}[b]
\centering \includegraphics[width=7cm]{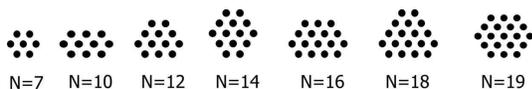} \caption{Sketch of toroid cross sections
for the first seven magic numbers.  Note that while all have high
degrees of symmetry, the $N=7$ and $N=19$ cross sections correspond
to perfect hexagons, and we dub these winding numbers
\emph{supermagic}. } \label{hexagons}
\end{figure}

The packing of filaments in a hexagonal crystal brings about particularly stable toroids for certain winding numbers, as noted previously in Refs.\ \cite{pereira2000,Schnurr}. This stability can be attributed to a high degree of hexagonal symmetry and the resulting low surface energy. From Table~\ref{alphatable} we can see that the difference between subsequent coordination numbers $\alpha_N$ is either 0 or 1 (for $N>2$). For instance, in the case of the 5-torus bundle shown on the right of Fig.\ \ref{bundle}, the addition of a sixth filament to the bundle can satisfy no more than two bonds, resulting in no fewer than four additional unsatisfied bonds and an increase in the coordination number by one. If this sixth filament is added just above the filament on the right, then the addition of a seventh filament directly above the center results in no increase in the coordination number, since three bonds can be satisfied. Here, the result is a symmetric, compact cross section that we refer to as a \emph{filled shell}. In such cases, where $\alpha_{N}=\alpha_{N-1}$, we refer to $N$ as a \emph{magic number}, following Refs.\ \cite{pereira2000,Schnurr}.  Figure~\ref{hexagons} shows toroid cross sections for the first seven magic numbers.  The magic numbers, up to $N=24$, are as
follows: $N=7,10,12,14,16,18,19,21,23,24$. If, instead, $\alpha_{N}=\alpha_{N-1}+1$, then the smaller $N-1$ torus is favored by both the interaction energy, as well as the bending energy. 

As winding number goes up, we have a higher density of magic
numbers.  This can be understood by the increase in edge vs corner filaments in the 
filled shells for large $N$. For instance, the $N=16$ structure in Fig.\ \ref{hexagons} can be obtained by removing three edge filaments from the $N=19$ structure. All but the last one of these filaments 
satisfy three bonds, corresponding to no change in $\alpha$, while the final corner filament 
satisfies only two bonds, corresponding to a reduction in $\alpha$. Thus, we find sequences of increasing length of successive magic numbers, although equilibrium tori are only found for the largest $N$
in each sequence. 

These sequences can be identified as follows. Certain winding numbers correspond to perfect hexagons, such as $N=7$ and $N=19$ in Fig.~\ref{hexagons}, which we call \emph{supermagic} numbers, as in Ref.\ \cite{Schnurr}. Since these hexagons consist of 6 equilateral triangles of length $k=1, 2, 3, \ldots$ filaments on each side plus one filament in the middle of the hexagon, the supermagic sequence is given by $N=3k(k+1)+1$. The coordination number in this case is 
$\alpha_N=3(2k+1)$. Thus, the difference between successive supermagic $\alpha$ is 6, although the difference in $N$ is $6k$, where $k$ corresponds to the larger $N$. Hence, the length of each sequence of $N$ with the same coordination number is $k$, on average. In fact, as suggested by Table 1, the actual lengths of the sequences of successive $N$ with equal coordination number are given by $k-1,k,k,k,k,k+1$. 

As noted before, semiflexible polymer condensates generically form rings or toroids from the competition
between their tendency to minimize surface area due to short-range attractive forces and their tendency to straighten out due to their substantial bending stiffness.  Balancing these two effects, i.e., setting 
$\kappa / L \sim \gamma L$, lets us define a natural length scale for our problem, which we call the condensation length, $L_c=\sqrt{\kappa/ \gamma}$. Physically, this length is the approximate length scale  at which we expect condensation to occur. Below this length DNA will rarely self-intersect and thus rarely condense, while above it a DNA filament will self-intersect many times and thus form collapsed, intermediate structures.

We can also define an analogous energy scale, the condensation
energy, $U_c=\sqrt{\kappa \gamma}$.  Given these scales, we can
present our conformational energies in dimensionless units, with
physical values of length and energies normalized by their
condensation values: $F_N \equiv U_N / U_c$, where $U_N$ is the
conformational energy of an $N$ torus, and $\lambda \equiv L / L_c$.
The presentation of our results in dimensionless units clarifies the
relevant parameters in our theory, namely the stiffness constant
$\kappa$ and the interaction parameter $\gamma$.

Combining our expressions for bending and surface energy and
normalizing by the condensation energy gives us the dimensionless
free energy for an $N$ torus

\begin{equation}
F_N =\frac{2\pi^2 N^2}{\lambda} + \lambda
(\frac{\alpha_N}{N}- \alpha_1) \label{prelimenergy}
\end{equation}

\subsection{Equilibrium Torus States}

Plotting Eq.~(\ref{prelimenergy}) for various $N$ gives a family of
curves which tend to negative infinity.  While there is no definite
global minimum in the free energy for all lengths, for a specific
reduced length $\lambda$, there is an associated optimal winding number
$N({\lambda})$, which corresponds to the equilibrium state at that length. For $\lambda
< 4 \pi $, the rod, $N({\lambda}) = 0$, is the equilibrium state.
From $N=2$ to $N=7$, every state is an equilibrium solution except
for $N=6$, which we expect from our discussion of magic numbers.
(These results differ somewhat from Ref.\ \cite{Schnurr}, since we focus on only tori of integer winding number.) 

Above $N=7$, all equilibrium states are magic number states, though
not all magic number states are equilibrium states. Instead, only the largest in each sequence of consecutive magic numbers winding numbers correspond to equilibrium tori in the absence of tension. This can be understood as follows. For $N>12$, the ratio $\alpha_N/N$, and therefore the interaction energy, has local minima at each value of $N$ such that both $\alpha_N=\alpha_{N-1}$ and 
$\alpha_{N+1}=\alpha_{N}+1$. This forms a sequence of winding numbers $N=14, 16, 19, 24\ldots$ for which particularly stable tori are expected. As can be seen in Tab.\ \ref{alphatable}, these all correspond to magic numbers. Including the effect of bending energy, which always favors smaller winding numbers, consistent with Ref.\ \cite{Schnurr}, we find stable tori for this sequence, as well as for the other magic numbers $N=7, 10, 12$. 

\begin{figure}[h]
\centering \includegraphics[width=8cm]{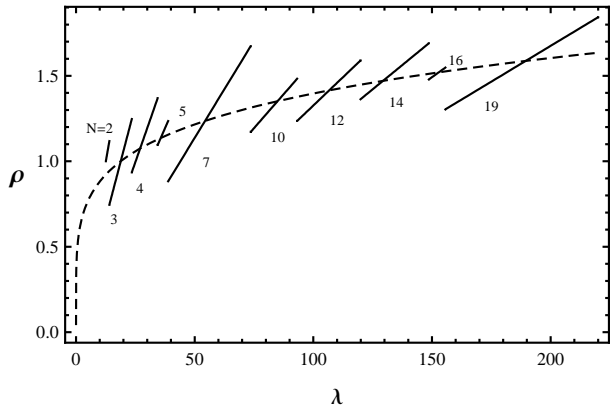}\caption{Plot of reduced radius vs reduced length for equilibrium tori up to $N=19$. The dashed line indicates the prediction based on an approximation valid for large $N$~\cite{Ubbink1995,Schnurr}.}
\label{radvlength}
\end{figure}

It has been noted before~\cite{pereira2000,Schnurr} that the equilibrium radii of the 
tori do not increase monotonically as a function of
reduced length. In fact, as reduced length is increased, the radii
of subsequent equilibrium tori states are marked by discontinuous
jumps.  Fig.~\ref{radvlength} shows the reduced radius, $\rho$, of
the equilibrium states as a function of reduced length, up to
$N=19$. The discrete transitions between the radii of different
winding number is again an effect of the hexagonal packing, which
creates islands of stability for certain winding numbers.  An
equilibrium toroid grows in radius until it reaches a contour length
at which the next equilibrium winding number is favored, at which
point it transitions to this state.  The extra length needed for the
additional loops of the higher winding number torus drives the toroid
to take on a smaller radius for the same contour length. For comparison, we also show as the dashed line in Fig.\ \ref{radvlength} the prediction based on a continuous approximation valid for large $N$ \cite{Ubbink1995,Schnurr}.

\subsection{Tori Under Tension}

The dimensionless free energy expression given in
Eq.~(\ref{prelimenergy}) is a measure of the internal energy of the
toroid.  At zero temperature this is analogous to the Helmholtz free
energy of our system. This free energy depends on the reduced
length of the filament, or the volume of our essentially 1D
system.  Since we also want to consider the effect of tension, which
as a force variable plays the role of pressure in a classical
thermodynamics analogy, we perform a Legendre transform on the
internal energy of the toroid to get our energy expression in terms
of force
\begin{equation}
E_N = U_N - \frac{dU_N}{dL}L =U_N + fL \label{legendre},
\end{equation}
where we equate $- \frac{dU_N}{dL} = f$.  Here, $E_N$ is analogous to the Gibbs free energy, 
where tension is the control variable.

The dimensionless form of the Gibbs free energy is obtained again by
normalizing Eq.~(\ref{legendre}) by $U_c$
\begin{equation}
G_N = \frac{2\pi^2 N^2}{\lambda} + \lambda (\frac{\alpha_N}{N}-
\alpha_1) + \tau \lambda \label{tensionenergy},
\end{equation}
where $\tau = {f}/{\gamma}$ is the dimensionless tension.
Equation~(\ref{tensionenergy}) represents the dimensionless result for
the toroid's conformational energy under tension that we will
generally be referring to when we discuss toroidal energy.

\section{Review of Experiments}\label{secExpts}

Over the past few years, a number of single-molecule %force spectroscopy 
experiments have probed the mechanics of DNA condensation under tension \cite{Baumann2000,Murayama2003,Besteman2007,Todd2008,Husale2008}.  
Generally, in these experiments a single DNA molecule is stretched and relaxed using optical or magnetic tweezers. Under 
conditions appropriate for condensation, most of these studies reported a nearly constant force plateau of several pN for DNA extensions lower than $\sim$85\% of the full contour length. While exact numbers differ between experiments, similar qualitative behavior was observed for a wide range of condensing agents and concentrations.

It has been generally believed that the force plateau regime consists of a continuous unraveling of DNA under tension. In order to test this hypothesis, we measured the force-extension relationship of a single condensed DNA molecule with high resolution using optical tweezers. Lambda-DNA (48502 bp) was attached on both ends to two optically trapped polystyrene beads \cite{BramThesis}  
and allowed to condense in the presence of 1 mM spermine4+. The condensed DNA was subsequently unraveled by displacing one of the beads. In these experiments, as shown in Fig.\ \ref{exptfig}, the force was allowed to vary freely, while the extension was varied by displacing the trap. We observed a roughly constant force plateau that was consistent with previous experiments. In detail, however, we found that this plateau consists of a sawtooth-like pattern, which suggests a step-like unraveling of the DNA under increasing extension. With the model above, we can account for these discrete steps in terms of jumps between toroid states with different winding numbers. In addition, this model can also account for the nearly constant force plateau. 

\section{Results of the Model}\label{theResults}

We find, for finite filament length and zero tension, that tori
states are, indeed, the equilibrium states within our model, as
discussed above. The filament length dictates which torus state has
the lowest free energy. For a given filament length $\lambda$
and winding number $N$, there exists a finite tension $\tau_{crit}$
at which the $N$ torus begins to unravel. For tension
$\tau<(\alpha_1-\alpha_N/N)$, the free energy $G_N$ strictly
decreases with increasing $\lambda$, resulting in a stable or
metastable state in which the torus incorporates the full polymer
length. Even before this point, however, the free energy
$G_N(\tau;\lambda)$ may become greater than zero, indicating that
the thermodynamically stable state is the extended polymer
conformation. In this case, the tori are actually metastable 
states.

With increasing tension $\tau>(\alpha_1-\alpha_N/N)$, a local
minimum in $G_N$ develops for $\lambda$ less than the full polymer
length. Physically, this corresponds to a mechanical metastable state,
or local energy minimum, in
which a torus coexists with a segment of unwound, straight polymer
under tension. The condition for this to occur is
\begin{equation}
\tau>-\frac{\partial F_N}{\partial\lambda},
\end{equation}
where the derivative on the right is evaluated at the full polymer
length. As the tension increases, more filament is pulled out of or
unwound from the torus, which then shrinks in size.  (Here, and
throughout, we assume that the torus is able to relax by internal
relative sliding of polymer.) As the tension increases and the torus
shrinks in size, the increased bending energy eventually results in
destabilization of the $N$-torus relative to tori of smaller winding
number. We identify below a series of transitions under tension to
tori of smaller winding number.

\subsection{Transitions}\label{sec:Transitions}

As tension is increased from zero, the slope of the free energy curve
for each $N$ increases.  
With increasing tension, the $N({\lambda})$ torus initially in equilibrium in 
the absence of tension for a given total length $\lambda$ remains the lowest energy state until a
critical tension is reached.  At this critical tension,
$\tau_{crit}$, the asymptotic slope of the free energy becomes 
zero for large $\lambda$. For tensions above this critical tension,
the $N({\lambda})$ torus becomes only metastable, as the ($N=0$) rod is now
energetically more favorable (with a free energy of zero), and is
thus the equilibrium state of the system.  As the tension continues to
increase, a local minimum of the free energy develops and starts to shift to lower values of $\lambda$.  Once this minimum shifts to values of $\lambda$ less than the 
full length of the DNA strand, then the metastable state consists of a compact $N$ torus, with a segment of
filament pulled out of the torus---i.e., the torus begins to unravel. 
With increasing tension, as more of the filament is pulled out of the
torus, larger $N$ tori become unstable to tori with smaller winding numbers. This 
unraveling process is sequential, with transitions to smaller and 
smaller values of $N$ as the tension increases. 

\begin{figure}[b]
\centering \includegraphics[width=8cm]{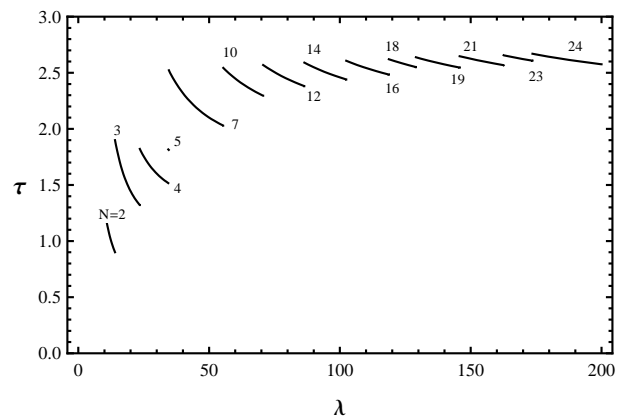}\caption{Tension vs reduced length
for preferred tori states from $N=2$ to $N=24$, only plotted at
tensions and lengths where we expect them. We note that only magic
winding numbers appear for $N>7$ and that a rough force plateau
appears at $\tau \approx 2.5$.  Transitions between winding numbers
are characterized by discontinuous jumps in tension and length.}
\label{goldenplot}
\end{figure}

With an eye toward addressing 
the experiments in Refs.\ \cite{BramThesis,Bram}, we consider a process in which
the reduced length $\lambda$ in the torus is controlled and 
slowly reduced, while the tension is allowed to 
vary. As the $N$ torus is slowly unravelled, the 
tension $\tau$ will increase. However, as $\tau$ increases, we 
expect at some point to develop a local minimum of Eq.\ (\ref{tensionenergy}) for 
$N-1$ that becomes less than or equal to that of the (metastable) $N$ toroid. 
Once a transition to the $N-1$ toroid occurs, if the length $\lambda$ is fixed, then the 
tension will fall as a new (metastable) $N-1$ toroid is formed at $\lambda$. This describes most
of the unravelling transitions, at least for large $N$, where $\alpha_{N-1}=\alpha_{N}$. 
However, given the discrete nature of the  
coordination number $\alpha_N$, it can happen that the $N-1$ state is itself unstable to the $N-2$ state
at $\lambda$. This occurs for $N>9$ whenever $\alpha_{N-2}<\alpha_{N-1}$, i.e., when $N-2$ is a magic number. 
Thus, the $N-1$ state can be expected to make a transition to the $N-2$ state: 
in the limit of a slow unravelling of the torus, the $N-1$ state is skipped. 
The resulting sequence of states and corresponding tensions vs 
reduced polymer length $\lambda$ is shown in Fig.~\ref{goldenplot}. Note
that the extension of polymer pulled out of the torus varies inversely in 
$\lambda$, meaning that metastable branch corresponds to a stable force-extension
relation, in which force increases as more polymer is pulled out of the torus. 
We find, interestingly, for $N$ greater than about 7, a nearly constant force \emph{plateau} at $\tau
\approx 2.5$. 
This is consistent with several recent experimental observations of a nearly constant force for tori under tension \cite{Baumann2000,Murayama2003,BramThesis,Bram}, as illustrated in Fig.\ \ref{exptfig}. 

\begin{figure}[h]
\centering \includegraphics[width=8cm]{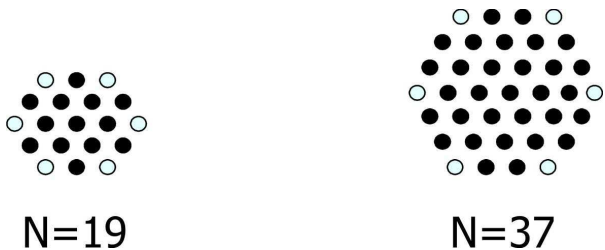} \caption{Cross sections of $N=19$ and $N=37$
tori with the lighter circles corresponding to corner filaments. As
winding number increases the number of corner filaments always
remains 6, so the overall density of corner filaments goes down for
large $N$ tori.} \label{bonds}
\end{figure}

The force plateau can be interpreted as the average force needed to
pull off a loop from the torus. Since the number of bonds broken
when pulling off a loop is either 2 or 3, depending on whether or
not the toroid has a magic winding number, it's not surprising that
we should see a force plateau at $\tau \approx 2.5$. As we go to
higher and higher winding number, however, the force plateau asymptotes to 3.
We can see why this is by considering the average number of bonds
per filament as a function of winding number. In Fig.\ \ref{bonds}, the most 
weakly bound filaments are the corner ones, which satisfy only three bonds. The remaining filaments along one edge adjacent to this corner can also be removed at the cost of just three bonds each, until the final corner filament along that edge is reached, the removal of which involves the breaking of just two bonds to form the magic number bundle with the next lowest coordination number ($N=16$ and $N=33$ for the bundles in Fig.\ \ref{bonds}).  With increasing winding
number, the fraction of filaments forming three bonds increases, and $\tau\rightarrow3$,
although this convergence is slow. 

\subsection{Large $N$ Behavior}
For large winding number $N$, the number of exposed, unsatisfied bonds at the perimeter of a 
torus cross section increases as $\sqrt{N}$, so that $\alpha_N \sim \sqrt{N}$. The prefactor
here is easy to calculate for perfect hexagons, as illustrated in Fig.\ \ref{bonds}. As noted in Sec.\ \ref{theModel}, these occur for $N=3k(k+1)+1$, where $k=1,2,3,\ldots$, for which the coordination number is  
\begin{equation}
\alpha_N=3(2k+1)=\sqrt{3\left(4N-1\right)}.\label{LargeNAlpha}
\end{equation} 
This actually represents a lower bound on $\alpha_N$, in general, since less symmetric 
cross sections have increased surface-to-volume or circumference-to-area ratios. 
As we are interested in the large $N$ behavior, we will approximate 
\begin{equation}
\alpha_N\simeq2\sqrt{3N},\label{LargeNAlpha}
\end{equation} 
as in Ref.\ \cite{Schnurr}.

\begin{figure} \centering \includegraphics[width=8cm]{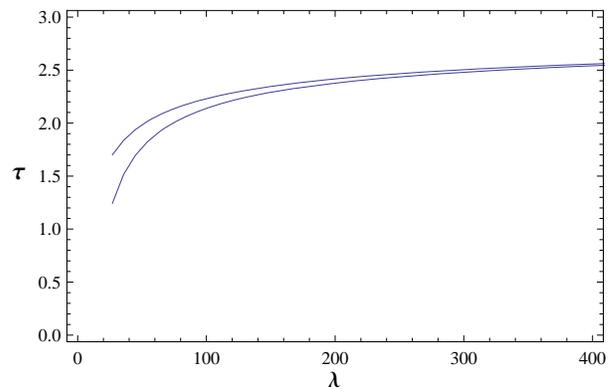}
\caption{The range of reduced tensions $\tau$ predicted for the metastable states, beginning with the $4\rightarrow3$ transition, using the large-$N$ approximation of Eq.\ (\ref{LargeNAlpha}).} \label{TauLargeN}
\end{figure}

\begin{figure}[b] 
\centering \includegraphics[width=8cm]{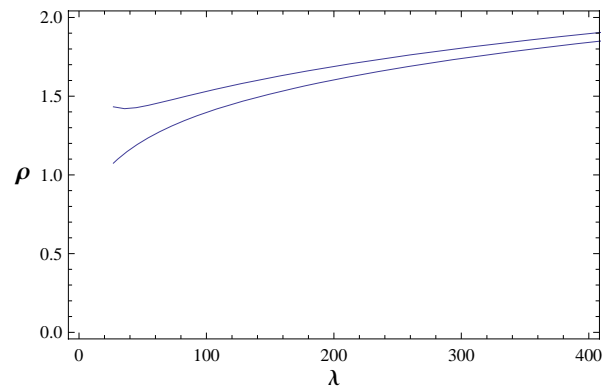}
\caption{The range of reduced torus radii $\rho$ predicted for metastable states, beginning with the $4\rightarrow3$ transition, using the large-$N$ approximation of Eq.\ (\ref{LargeNAlpha}).} \label{RhoLargeN}
\end{figure}

We use this large $N$ approximation to determine the metastable states and transitions between them, as we have done 
in the previous section. Specifically, we consider fixed but decreasing condensed length $\lambda$. 
In Figs.\ \ref{TauLargeN} and\ \ref{RhoLargeN}, we indicate the predicted sequence of 
metastable states and corresponding values of tension $\tau$ and toroid size $\rho$ in reduced units. 
We find many of the same qualitative features in this large $N$ approximation as were found 
in the previous section. In particular, we find an apparent force plateau, much 
as in Fig.\ \ref{goldenplot}. Although both models predict the same asymptotic convergence of 
the metastable tension $\tau\rightarrow3$ for large $\lambda$, where $N$ is also large, we see 
that the approximate value of $\alpha_N$ from Eq.~(\ref{LargeNAlpha}) yields a consistently smaller value of the tension
than for the discrete model in Sec.\ \ref{sec:Transitions}. This can be understood as follows. 
In the discrete model, we account for the hexagonal packing of filaments, which results in 
a sequence of $\alpha_N$ in which there are discrete jumps in $\alpha_N$ (at non-magic numbers), between which 
ranges of constant $\alpha_N$ are found. For transitions from $N$ to $N-1$ toroid states, the
$N-1$ state is thus destabilized in the discrete model as compared to the large $N$ approximation. when $\alpha_N=\alpha_{N-1}$.
This enhanced relative stability of the $N$ toroid means that the tension $\tau$ is larger at the 
transition. For the same reasons, the corresponding toroid sizes are smaller in the discrete model 
than in the large $N$ model. Nevertheless, the general features, and especially the force plateau, are 
seen for both models. 

\subsection{Energy Barriers}\label{SecBarriers}

\begin{figure}[h] 
\centering \includegraphics[width=8cm]{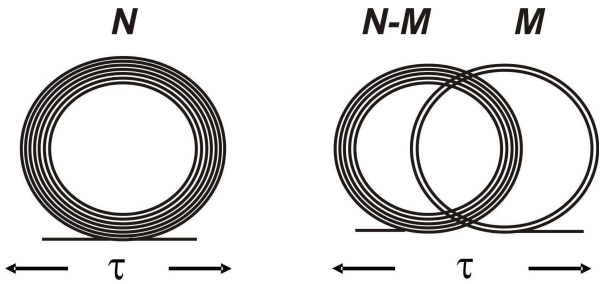}\caption{Sketch of $N$, $N-M$ and $M$ tori. Physically we
expect the energy barrier between these to be dependent only on
surface energy difference, as the bending energy and tension are
unchanged.} \label{energysketch}
\end{figure}

We estimate the energy barrier between $N$ and $N-M$ tori ($M<N$)
first by calculating the energy difference between an $N$ torus and $N-M$
+ $M$ tori (see Fig.~\ref{energysketch}). This is not to suggest that 
Fig.\ \ref{energysketch} represents the actual reaction pathway for 
the transition between tori: calculating the energy in this way, 
by considering the separation of full loops from the original torus, can only 
provide an upper bound on the transition energy. From our model, and 
as suggested by experiment, we expect that toroids make step-wise
jumps between different winding numbers.

Intuitively, we expect the transition to depend primarily on the difference in
surface area exposed to solvent, and not on the bending energy, since
the torus radii vary only weakly with $N$. In what follows, we shall
assume that the radii are constant.  
We can see from Fig.~\ref{energysketch} that the following
relations hold, since the total amount of reduced length,
$\lambda_N$, is unchanging and all the loop radii are equal:
$\lambda_N=\lambda_{N-M} + \lambda_M$, where
$\lambda_{N-M}=\frac{N-M}{N}\lambda_{N}$ and
$\lambda_{M}=\frac{M}{N}\lambda_{N}$.  Here, $\lambda_N$ is the
length in the $N$ torus, $\lambda_{N-M}$ the length in the $N-M$
torus, and $\lambda_{M}$  the length in the $M$ torus.

\begin{figure}[b]
\centering \includegraphics[width=8cm]{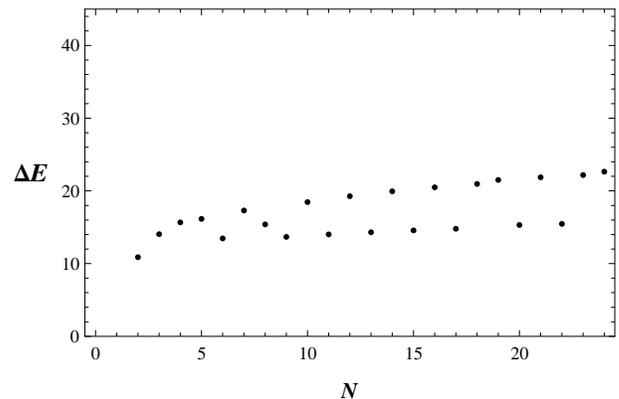}\caption{Estimated energy
barriers (dimensionless) vs torus winding number, using Eq.\ (\ref{energybarseqn}), for all $N$ to
$N-1$ transitions between $N=2$ and $N=24$. The energy barriers
are between $N$ and $N-1$ tori, e.g., the energy barrier plotted at $N=12$ is the
energy barrier between $N=12$ and $N=11$.} \label{energybarriers}
\end{figure}

With these identifications we can straightforwardly calculate the
energy difference between the $N$ torus and the $N-M$ + $M$ tori,
from the energy expression given by Eq.~(\ref{tensionenergy})
\begin{eqnarray}
\Delta E_{N,N-M}&& \!\!\!\!\!\!= \frac{2\pi^2 (N-M)^2}{\lambda_{N-M}} +
\lambda_{N-M} (\frac{\alpha_{N-M}}{N-M}- \alpha_1)\nonumber\\ 
&&\!\!\!\!\!\!\!\!\!\!\! +
\tau \lambda_{N-M}  + \frac{2\pi^2 M^2}{\lambda_M} +
\lambda_{M} (\frac{\alpha_M}{M}- \alpha_1) + \tau \lambda_M  \nonumber\\
\nonumber
&&\!\!\!\!\!\!\!\!\!\!\! - \frac{2\pi^2 N^2}{\lambda_N} - \lambda_N
(\frac{\alpha_N}{N}- \alpha_1) - \tau \lambda_N\nonumber\\ 
&&\!\!\!\!\!\!\!\!\!\!\! = \frac{\lambda_N}{N}(\alpha_{N-M} + \alpha_M -\alpha_N),
\label{energybarseqn}
\end{eqnarray}
which depends only on the surface parameters, as expected.  
We calculate $\lambda_N$ in the above equation as follows; starting from an $N$ torus we assume downward sequential transitions in winding number, as previously described, occurring for $\lambda_N$ where the metastable $N$ and $N-1$ states have equal free energy.  For unstable tori, we assume a direct transition to the subsequent winding number at the same value of $\lambda$. It is not our purpose to calculate realistic rates of such transitions. Instead, we focus on identifying the most relevant sequence of metastable torus states in a fixed-$\lambda$ ensemble, corresponding to experiments in which the total length is the control variable. Figure~\ref{energybarriers} shows the energy barriers between $N \rightarrow N-1$ transitions, from $N=24$ down. 
We find that transitions from $N$ to $N-M$ for $M>1$ have substantially higher energy barriers, and are thereby strongly suppressed. 

As noted above, it is unlikely that the transitions sketched in Fig.\ \ref{energysketch} represent the real reaction pathway between torus states. Nevertheless, our estimates suggest that transitions corresponding to a change in winding number by more than 1 are strongly suppressed. For $\Delta N=1$, however, it is also possible that DNA peels off continuously from the torus. This will, in general, induce additional bending as the DNA is pulled off the torus. (We assume twist is relaxed.) In fact, a bend of approximately 90 degrees is expected on simple mechanical grounds, since the force in the plateau region in Fig.\ \ref{goldenplot} is comparable to the total binding energy per unit length of filament, as discussed in Sec.\ \ref{sec:Transitions}. Under a force $f$, the radius of curvature $r$ for such a bend can be estimated by balancing the total bending energy $\pi\kappa/(4r)$ and the virtual work against the applied tension $(\pi/2-1)rf$. This yields an optimal radius of curvature $r=\sqrt{\pi\kappa/[f(2\pi-4)]}$, and a total energy for two such bends of $\sqrt{\tau\pi(2\pi-4)}\approx 4.2-4.6$ in reduced units for $\tau\approx 2.5-3$. This is about a factor of 2-3 smaller than the estimates from Eq.\ (\ref{energybarseqn}). 

Other effects may change the energy barriers as well, such as
next-nearest neighbor interactions.  As these interactions are
effectively attractive, they will favor more compact structures,
thus lowering the length needed for an $N$ torus (i.e. shifting
minima in the energy to lower values of $\lambda$).  This will lower
the energy barriers given above for nearest neighbor interactions,
but there will be an additional contribution to the energy barrier
from the attraction of next-nearest neighbors that must be overcome
to pull off a loop.  Without explicit inclusion of these effects in
our model it is unclear what net effect these interactions will have
on energy barriers. 

\section{Discussion}\label{theExperiment}

Of the two parameters in our model, the bending stiffness $\kappa$ is known to be approximately 
$\kappa = kT\ell_p\simeq 50kT\cdot$nm, where $\ell_p$ is the
persistence length of DNA.  The interaction
parameter $\gamma$ is expected to depend on the counter ions present 
in solution, and it is unfortunately not known.  In order to estimate $\gamma$, we use the experimentally 
observed plateau value of the force in Fig.\ \ref{exptfig}.  This is 
consistent with the plateau we find in
Fig.~\ref{goldenplot}.  
By matching the measured force plateau with our
(dimensionless) tension $\tau={f}/{\gamma}$ we 
estimate $\gamma \simeq 1.6 $pN.  This allows us to estimate
the condensation length, $L_c \simeq 11 $nm, and condensation energy,
$U_c \simeq 4.5 kT$.  This corresponds to energy barriers of order 
20$kT$, using the lower estimates at the end of Sec.\ \ref{SecBarriers} for
$\Delta N=1$ transitions. By contrast, we find substantially larger barriers of more than 50$kT$ for 
transitions as sketched in Fig.\ \ref{energysketch} for $\Delta N>1$. 

Figure~\ref{goldenplot} shows a typical
transition length (i.e., the difference in reduced length between an
$N$ torus and an $N-1$ torus at constant tension) of $\sim 10$,
suggesting typical torus loop sizes of $\sim 112$nm, a value larger than reported 
in Ref.\ \cite{Bram} by a factor of about 2-3. 
Given the simplicity of our model, being off by such a factor is
perhaps not so bad, as there are many effects that we have not taken
into account. One notable effect that we've neglected is
next-nearest neighbor interactions.  Such interactions will have the
same attractive tendency as nearest neighbor bonds, and thus will
tend to favor more compact structures, translating into smaller loop
sizes.

Another possible effect is stability of the toroids beyond the
transition regime. We have predicted the loop sizes from our model
assuming immediate transitions once in the transition regime; in
reality toroids may not transition immediately.  At higher tensions
there are smaller differences in reduced length between subsequent
toroid force-extension curves. If,
for example, we instead assume a force plateau at the
higher-than-expected value $\tau=3$ (which raises $L_c$ to 12nm),
and average over the same range of winding numbers, we get an
average step size of $\lambda \simeq 6.5$, translating into a loop
size of 78nm.

\begin{figure}\vspace{5mm}
\centering \includegraphics[width=8cm]{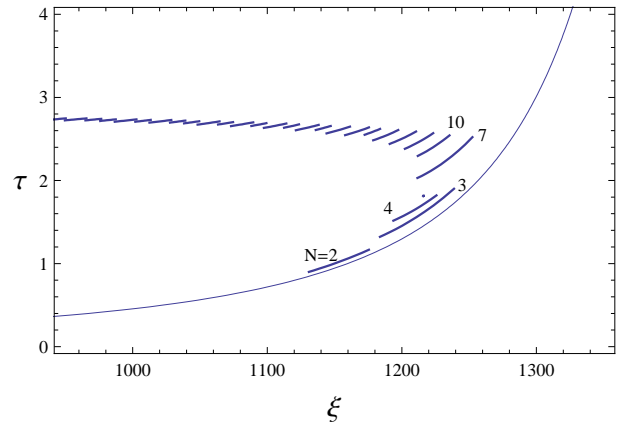}\caption{The predicted force (in units of $\gamma$) as a function
of the apparent extension $\xi$ (in units of $L_c$) of DNA toroids in series with freely fluctuating uncondensed DNA. Here, we have used $16.4\mu$m as the full contour length and $L_c=11$nm. A few of the individual force-extension curves are labeled by the corresponding winding numbers $N\leq 10$. The thin continuous curve represents the force-extension curve of the bare
DNA.} \label{FigCompare}
\end{figure}

One very important effect that must be taken into account in comparing with the 
measured extensions in the experiments is the finite extensibility of the uncondensed DNA strand, which 
is expected to be well approximated by a worm-like chain \cite{Marko,BramThesis,Bram}. We can account for this
by considering a thermally fluctuating filament in series with the torus. The length of this filament is 
equal to the full contour length $L_0$ of the DNA in the absence of
condensation, minus the length $L$ contained in the torus. 
In dimensionless form, the extension $\xi=x/L_c$ of the free DNA in series with the toroid is approximately given by 
\begin{equation}
\xi=\left(\Lambda-\lambda\right)\left(1-\frac{kT}{U_c\sqrt{\tau}}\right),
\end{equation}
where the full contour length of uncondensed DNA is given by $\Lambda L_c$, of which $\lambda L_c$ is contained in the toroid. 
In Fig.\ \ref{FigCompare} we show the combined force-extension curves of the 
the various toroid states in Fig.\ \ref{goldenplot}, corresponding to $U_c \simeq 4.5 kT$ and the 
full contour length of 16.4$\mu$m in Refs.\ \cite{BramThesis,Bram}.
Interestingly, this model predicts multiple metastable states with winding numbers in the range of approximately 2-10 near the transition to the fully unraveled toroid. This may explain the significant hysteresis reported for forward (extension) and reverse force-extension measurements near the transition between toroid and fully extended DNA \cite{BramThesis,Bram}. As the fully extended state is allowed to condense when reducing the extension, the force drops and toroid states with small $N$ are expected to form at a smaller extension than required, e.g., for the transition from $\simeq 7$ to fully extended DNA. 

%\section{Discussion}

The simple model developed here is able to capture a number of features observed in the experiments. We 
find theoretically that toroid unraveling under tension occurs via a series of discrete transitions, as 
observed in Refs.\ \cite{BramThesis,Bram}. In addition, this model provides an explanation for the 
fact the approximately constant force plateau for DNA condensations under tension, as reported in a number of experiments
\cite{Baumann2000,Murayama2003,Besteman2007,BramThesis,Bram}.    

%\begin{acknowledgments}
 
%\end{acknowledgments}

\end{document}